# PHASE TRANSITION IN SIMPLEST PLASMA MODELS

**I.L. Iosilevski, A.Yu. Chigvintsev**

*Moscow Institute of Physics and Technology*
(Dolgoprudny, Moscow Region, 141700, USSR)

The well-known simplest plasma model One-Component Plasma (OCP) is the system of free moving charges of the same sign in the uniform compensating background of opposite sign. This model is studied carefully nowadays [1÷3]. It should be emphasized that this model is not the single one but it is the family of models, in fact. The difference may be in the additional short-range interaction and in the type of statistics, but the main subject for the following discussion is the difference in the nature and the thermodynamic role of background.

In the ordinary version of the model OCP with the rigid background - the volume variations are not defined. The system can not collapse or explode spontaneously and so the well-known negativity [1] of formally defined pressure and compressibility does not signify the thermodynamic instability. The only phase transition – crystallization occurs without change of density in this version of OCP.

Now the main subject of our interest is another variant of this model. It is OCP with the compressible but still uniform background. It means that the background does not screen the moving charges individually but does it in average only. More precisely, this version of model may be defined by means of two variants of definitions:

**A)** Through the overall electroneutral Grand canonical ensemble [4]

**B)** Through the density functional with the strong gradient correction at the Helmholtz free energy of background so that this term would tend to infinity ($A \to \infty$) after the thermodynamic limit ($N \to \infty$, $N/V = n = const$)

$$F_{\text{OCP+background}}[n_{\text{ocp}}(\mathbf{x}); n_b(\mathbf{y})] = F_{\text{OCP}} + F_{\text{background}}^{local} + A \int_V [\nabla n_b(\mathbf{x})]^2 d\mathbf{x} \qquad (1)$$

(The contribution of Coulomb interaction is included in (1) into the "OCP"-term).

It is known that the crystallization occurs with the small density variation in this model [5]. But the main statement of this report is the appearance of a new first-order phase transition [6], with the properties strongly depending on the precise definition of a background thermodynamic contribution. It should be noted that the existence of this phase transition may be proved independently of a calculation of phase transition parameters. The simple arguments [7] with the use of the Gibbs-Bogolubov inequality, dimensionless analysis and the well known lower bound for the pressure correction of OCP [4] tend us to the form of equation of state of this model in the high density limit ($n \to \infty$).

$$P = P_{background} + P_{background}^{ideal} - n^{4/3} const \qquad (2)$$

Two conclusions may be done:

1. The OCP of classical point charges in the Boltzmann ideal gas uniform background is thermodynamically unstable against a collapse at any temperature and density. This statement is in agreement, firstly, with the result of Lieb and Narnhofer [4] on a divergence of the OCP overall electroneutral grand canonical ensemble, and secondly, with the conclusion of Dyson [8] that thermodynamic stability of Coulomb system requires at least one sort of particles to be fermions.

2. In the case of ideal Fermi-gas background the phase transition of gas-liquid type with upper critical point appears at the sufficiently low temperature or sufficiently large value of charge.

**Classical Point Charges in the Ideal Fermi-Gas Background**

The equation of state of both these subsystems (OCP and background) are known almost exactly, so we can carry out direct calculations of the parameters of this phase transition. In this calculations we have used the analytical fits from [1, 5] (OCP) and [9, 10] (ideal Fermi-gas). The results are presented in Tab.I and Figs.1÷3 in standard notations.

$$\Gamma \equiv (4\pi n/3)^{1/3} (Z^2 e^2/kT) \quad \Lambda_e^2 \equiv 2\pi\hbar^2/m_e kT \quad r_S^{-3} \equiv 4\pi n_e a_B^3/3 \quad a_B \equiv \hbar^2/m_e e^2$$

|     | $Z_i$ | 1 | 2 | 3 | 10 | 30 | 100 | 1000 |
|---|---|---|---|---|---|---|---|---|
| I | $T_c$ (Ry) | 0.0393 | 0.152 | 0.314 | 2.17 | 10.6 | 55.8 | 1232 |
|   | $\Gamma_c$ | 8.86 | 12.6 | 16.3 | 42.7 | 117 | 376 | 3690 |
|   | $(r_S)_c$ | 5.73 | 3.32 | 2.43 | 1.00 | 0.467 | 0.206 | 0.044 |
|   | $(n_e\lambda_e^3)_c$ | 7.22 | 4.90 | 4.22 | 3.30 | 3.03 | 2.94 | 2.89 |
|   | $\{P/(n_i + n_e)kT\}_c$ | 0.125 | 0.129 | 0.130 | 0.130 | 0.129 | 0.128 | 127 |
| II | $T_{tr}/T_c$ | 0.113 | 0.148 | 0.177 | 0.417 | 9.37 | 41.35 | - |
|   | $(\Delta n/n)_{tr}$ | 0.019 | 0.019 | 0.020 | 0.021 | 0.069 | - | - |
| III | $(r_S)_{binodal\ (liq.)}$ | 2.47 | 1.55 | 1.19 | 0.531 | 0.255 | 0.115 | 0.025 |
|   | $(r_S)_{spinodal\ (liq.)}$ | 3.08 | 1.94 | 1.48 | 0.664 | 0.319 | 0.143 | 0.031 |
|   | $\Gamma_{spinodal\ (gas)}$ | 5.76 | 8.39 | 11.0 | 28.9 | 79.6 | 256 | 2518 |
|   | $\{\Delta H_f/N_Z\}_0$, Ry | 0.363 | 1.831 | 4.715 | 78.27 | 1016 | 16860 | 3.63+6 |

**Table I**. Parameters of phase transitions in OCP of classical point charges on the ideal Fermi-gas background for the different values of charge number Z: *I, II* – parameters of critical and triple points; *III* - the heat of sublimation, $\Delta H_f$ and parameters of condensed ($r_S$) and gaseous ($\Gamma$) binodals and spinodals in the limit $T \to 0$.

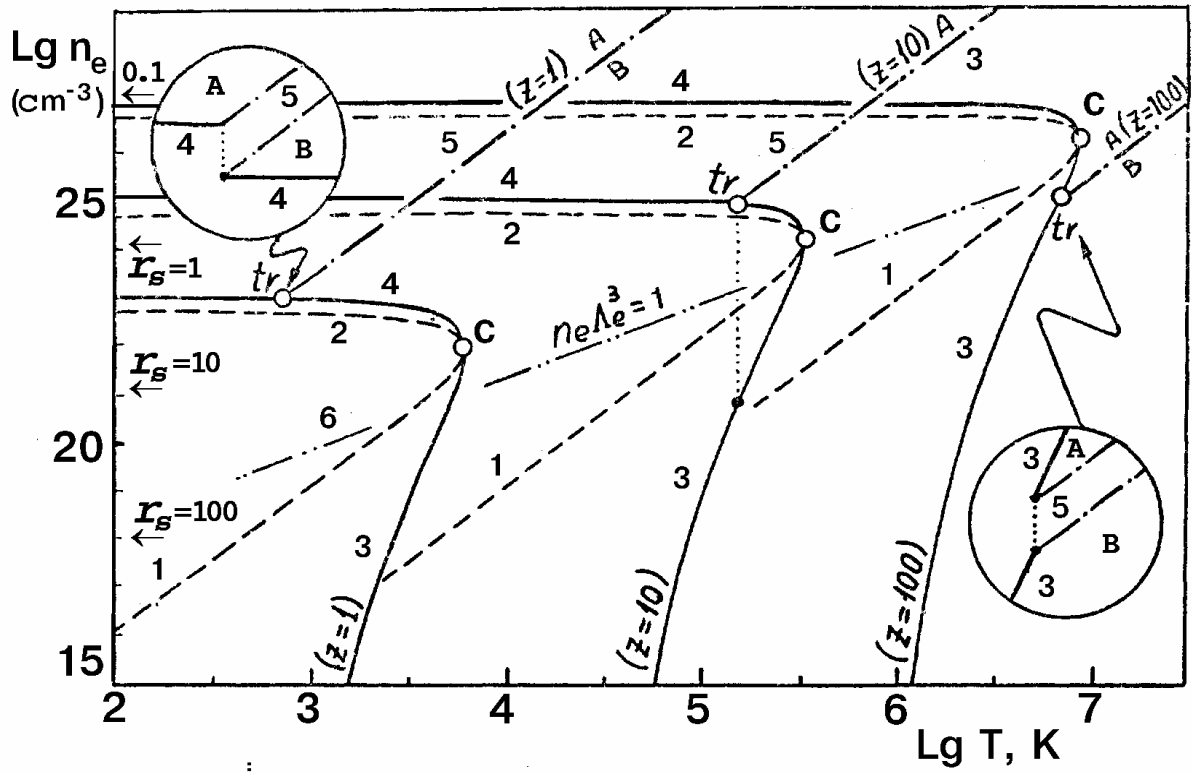

**Fig.1.** Phase diagram of OCP classical point charges in the ideal Fermi-gas background in temperature 5 background electron density coordinates for different values of charge number $Z$. Spinodals (*1, 2*) and binodals (two-phase coexistence curve (*3, 4*)) of condensed (*2, 4*) and gaseous(*1, 3*) phases, melting (*5*) of crystalline (**A**) to fluid (**B**) phases ($\Gamma \approx 178$), critical (*c*) and triple (*tr*) points, the position of electron degeneracy (*6*) and constant Brueckner parameter $r_S = 0.1$, 1, 10 and 67 (the last value corresponds to the 'cold' melting of electron Wigner crystal [11]) are remarked.

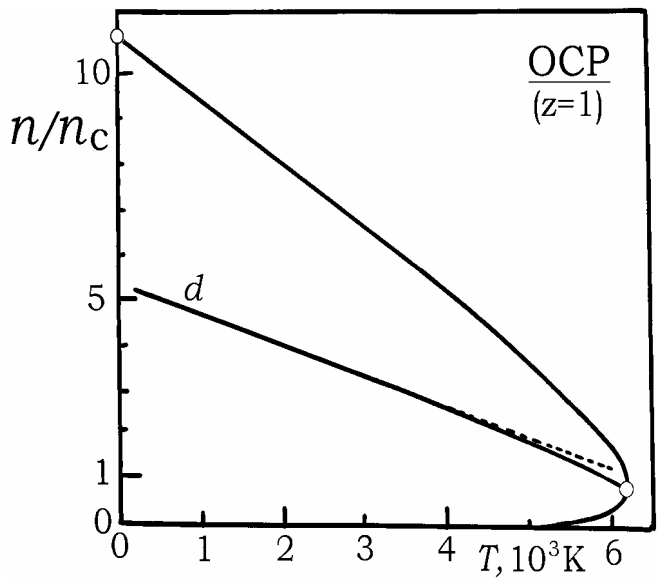
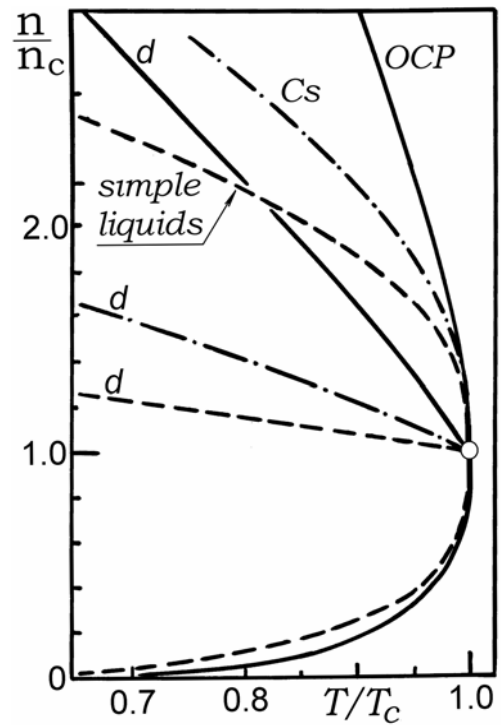

**Fig.2.(a,b)**. Density vs. temperature diagram of coexistence curve and its diameter:
$$d \equiv (n_L + n_G)/2$$

Some remarks should be done to these results:

1) The compressibility factor $Z_c \equiv \{P/(n_i + n_e)kT\}_c$ in critical point has almost the same rather low value for all charge numbers (*).

2) For $Z > Z^* \approx 46$ the melting curve (line $\Gamma = 178$) crosses the gaseous part of binodal and thus it forms specific picture of the phase diagram.

3) The density variation between crystalline and fluid phases in the triple point is fairly small (see Table I) when this point does not coincide with the critical point, in this latter case ($Z = Z^*$) the value of this variation increases remarkably. Besides that, this variation is much greater than the value estimated in [5] ($\Delta V/V \approx 0.0003$).

4) For present phase transition the deviation from the well-known semi-empirical rule of rectangular diameter is similar to those for alkali metals (Fig.2), but the slope of this diameter and a relation of normal to critical densities are rather great in comparison with the real substances. For example, for alkali metals $n_c/n_0 \cong 4 - 5$.

5) The critical point of this phase transition is not the genuine one because the density fluctuations in the vicinity of the critical point are absent in the model definition "**A**" and are suppressed by the gradient term in the definition "**B**". In the limit of the infinite gradient term the critical exponents must tend to their classical values. In the case of strong but finite gradient term in (1) the critical point position must slightly deviate from its limiting value.

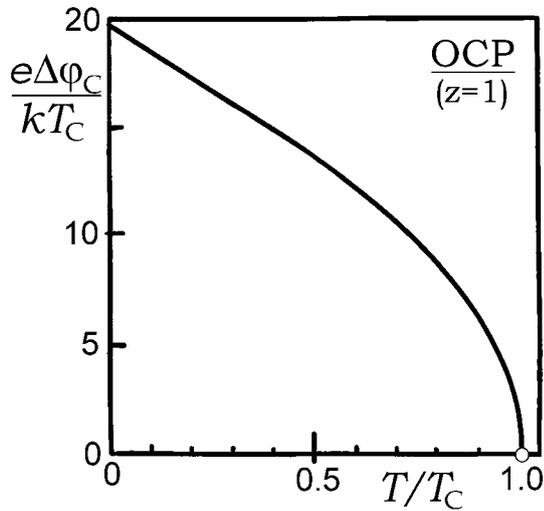

**Fig. 3** Potential drop between two phases: $\Delta\varphi \equiv \varphi_L - \varphi_G$

6) Remarkable feature of any non-uniformity in equilibrium system of Coulomb particles (in particular of inter-phase boundary) is a finite difference at the average electrostatic potential through this non-uniformity. The equality of two electrochemical potentials for every kind of Coulomb particles at both sides of non-uniformity leads to inter-phase potential drop depending on temperature (Fig.3)

$$\mu_i^{(1)} + Z_i e \varphi^{(1)} = \mu_i^{(2)} + Z_i e \varphi^{(2)} \Leftrightarrow$$

$$\varphi^{(2)} - \varphi^{(1)} = (Z_i e)^{-1} [\mu_i^{(1)} - \mu_i^{(2)}] \qquad (3)$$

The zero-temperature limit of this drop, $\Delta\varphi_0$ is the supplementary thermo-electrophysical constant of a substance. The high value of this drop for OCP model (see Fig.3) ought to be noted. Estimations for several metals [13] give the value $e\Delta\varphi_0/kT \sim 2\text{-}3$.

================================================

(*) It is quite close to the estimation, $Z_c = 0.145$ [12] with the use of equation of state of Van der-Waals type with Coulomb-like attractive term, $\Delta p \sim n^{4/3}$.

## Quantum Electron Gas

In this variant of OCP the moving charges are the electrons. When the background is rigid the melting curve changes its shape in the high density limit because of cold melting appearance. The estimations [14] give the Wigner crystallization inside the boundaries: $\Gamma \geq 178$; $r_s \geq 67$ [11] (classical and quantum melting) and $T \leq 10^{-5}$ Ry [14]. Transition to the electron gas in the compressible but uniform background changes this phase diagram essentially because of appearance of a vast phase transition of gas-liquid type. If we suppose that the parameters of this phase transition for Quantum Electron Gas are close to those of Classical OCP on the ideal Fermi-gas background, we should conclude that this new phase transition completely excludes the electron gas Wigner crystallization because of its melting curve placed deeply inside not only two-phase coexistence region, but just inside the spinodal curve (Fig.1). Therefore, at non-zero temperatures this crystal is absolutely unstable against the phase decomposition into two fluid phases of weakly coupled plasma: a dense phase and a rare one. For $T \approx 0$ this supposition was declared previously in [15].

## Double OCP model.

The background thermodynamic properties may be defined more meaningfully with the use of the variational principle of statistical mechanics. The choice of the *N*-particle distribution function in a multiplicative form as a product for nucleus and electron multipliers $\rho_N = \rho_{(+)} \cdot \rho_{(-)}$ is equivalent to the complete switching off individual correlations between two subsystems of charges while $(+)\leftrightarrow(+)$ and $(-)\leftrightarrow(-)$ correlations are completely permitted. So the nucleus and electrons turn for each other into the uniform and compressible compensating background. This combination of two inserted one into another OCP-models will be called here with the term "Double OCP". By switching off other kinds of dynamic correlations inside each sub-systems one can obtain the hierarchy of simplest plasma models with the decreasing free energy. Each of them presents the upper bound for a free energy of real plasma, the Double OCP being the best.

$$F_{(+-)} \leq F_{(+)}^{(OCP)} + F_{(-)}^{(OCP)} \left\{ \begin{array}{l} \leq F_{(+)}^{(HF)} + F_{(-)}^{(OCP)} \leq F_{(+)}^{(ideal)} + F_{(-)}^{(OCP)} \\ \leq F_{(+)}^{(OCP)} + F_{(-)}^{(HF)} \leq F_{(+)}^{(OCP)} + F_{(-)}^{(ideal)} \end{array} \right\} \leq F_{(+)}^{(ideal)} + F_{(-)}^{(ideal)} \quad (4)$$

Being applied to the real substances, the Double OCP model gives the phase diagram depending on the masses of positive charges. For heavy particles the phase diagram is similar to those in Fig.1. The main difference is additional cold melting in the nuclear subsystem in the high density limit (Fig.4). For the mass-symmetrical plasmas (electron-positron and electron-holes plasmas) the phase diagram is similar to this for electron gas with the only phase transition of gas-liquid type. The Double OCP model gives a simple estimation for the critical temperature of this phase transition. For example in electron-positron plasma $kT_C \approx 0.6$ eV.

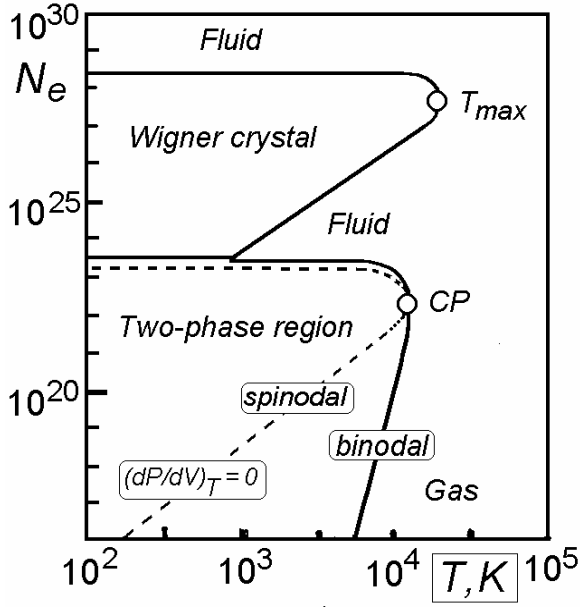

**Fig.4**. Phase diagram of the Double OCP model of electron-proton system. Present estimation with use of the data [14] and modified results of Fig.1.

**OCP Phase Transition and Anomalies in the Equilibrium Spatial Charge Distribution**

The discussed OCP phase transition seems to be practically unknown. Let us consider the situation where our knowledge of the properties of this phase transition may be useful. It is so in every case of a thermodynamically equilibrium spatial charge distribution in non-uniform plasma. An example is the quantum electron distribution in atomic cell. This problem may be formulated in terms of the Density Functional (for example [16]). The Free Energy Functional $F[n(W)]$ is minimized over the electron density $n(r)$ with the additional normalization condition

$$F = \min_{n(\cdot)} F[n(\cdot)] = Ze \int \varphi_{ext}(\bar{x}) \cdot n(\bar{x}) d\bar{x} + \frac{Z^2 e^2}{2} \int \frac{n(\bar{x}) \cdot n(\bar{y})}{|\bar{x} - \bar{y}|} d\bar{x} d\bar{y} + F^*[n(\cdot)], \qquad (5)$$

In the local approximation the "exchange-correlation-kinetic" term $F^*[n(\cdot)]$ is follow:

$$F^*[n(\cdot)] = \int f(n(\bar{x})) \cdot n(\bar{x}) d\bar{x} \; ; \qquad f(n) = \lim \left\{ \frac{F(N,V,T)}{N} \right\}_{(N \Rightarrow \infty; N/V = n)} \qquad (6)$$

When $F(N,V,T)$ is the free energy of Boltzmann or Fermi ideal gas, we deal with the well-known Poisson-Boltzmann or Thomas-Fermi approximations [16]. It ought to be emphasized that in frames of approximation (5)(6) any attempt to take into account electron correlations by using the exact expression for the free energy of electron gas, $F(N,V,T)$ will require the equation of state of OCP with the uniformly compressible background. As a result the discussed OCP phase transition appears in the case of sufficiently low density and temperature ($T < T^*$) in the form of discontinuity in a spatial charge distribution inside the cell. So the smooth Thomas-Fermi (or Poisson-Boltzmann in classical case) distribution will be subdivided into the condensed "drop" around the attractive center and diffuse "atmosphere" at the cell periphery. In terms of the OCP phase transition the boundary temperature $T^*$ is just equal to the OCP critical temperature (for electron gas it is $T_c \sim 1$ eV), and two values of local densities, $n_1^*$ and $n_2^*$ (see Fig.5) are equal to the densities of coexisting phases, condensed and gaseous, which are depending on temperature only. It ought to be noted especially that this phase decomposition occurs in the system of particles with entirely repulsive interaction.

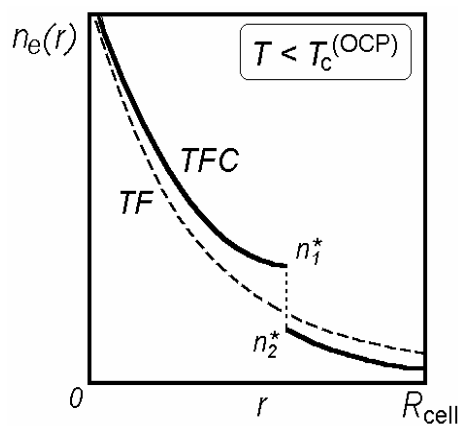
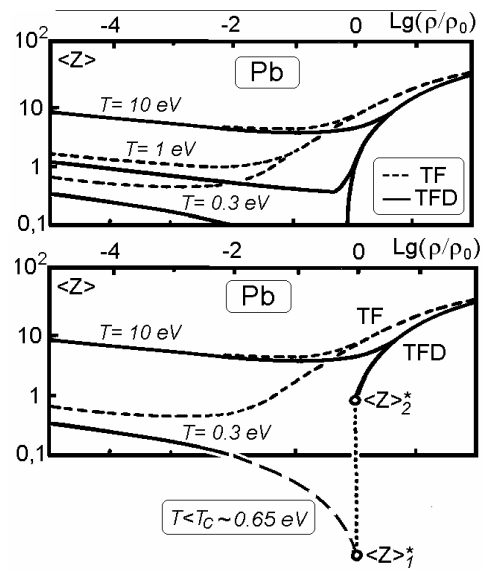

**Fig.5.** Electron gas profile in atomic cell at $0 < T < T_c^{OCP}$. Qualitative comparison of two variants of local approximations (5),(6): *1* - Thomas-Fermi approximation; *2* - additional taking into account of exchange and correlation corrections in frames of exact equation of state of strongly coupled OCP with uniformly compressible background.

**Fig.6.** "Ionization degree", <Z> of Pb atomic cell in TF and TFD approximations (Fig.2 from [19]). The curve under the main figure represents presently supposed profile of <Z> at low temperatures $0 < T < T_c^{OCP} \sim 1$ eV.

This example allows one to predict the appearance of similar phase stratifications in all the cases when the equation of state of strongly coupled OCP is being used to calculate the non-uniform charge distribution in local approximation (6). For example, it occurs in the case of equilibrium charges near charged hard wall or in the case of the background edge, the same-sign charges in macroscopic shell or cavity and etc. [7, 17].

It should be noted that it is not necessary to use in (6) the exact equation of state of strongly coupled OCP for the appearance of discussed discontinuity. For this it is sufficient to use any local attractive correction (exchange or correlation). Therefore this discontinuity must appear as a consequence of the Thomas-Fermi-Dirac approximations or other local modifications. It is really so, just the same discontinuities were shown in the figures (12÷15) at the famous book of P. Gombash [18]. Note that the density of a "gaseous" part of the profile in the Fig.5 is equal to zero in the case $T = 0$. For the case $T \neq 0$ the calculations of electron distribution in atomic cell in TFD-approximation have been made in [19]. The degree of ionization was defined in [19] to be proportional to the boundary electron density. One can conclude this quantity should be discontinuous for $T < T_c^{OCP} \sim 1.0$ eV. Unfortunately this part of results was omitted in the figures 1 and 2 of [19]. The supposed structure of omitted part of this curves at $T < T_c^{OCP}$ is shown schematically under the Fig.6. It ought to be noted that the value of this discontinuity must increase when $T \to 0$ so that $<Z^*>_1 \to 0$ and $<Z^*>_2 \to \text{const} \approx 0.035$ (this latter value corresponds to the boundary electron density at $r_S \approx 4.2$ and $P = 0$).

**Conclusions**

Next models from OCP-family mentioned above (with the compressible but uniform background) seem to be interesting for future study:

1. Double OCP plasmas (exact parameters calculation)
2. Binary ionic mixture (BIM). Interrelation of presently studied phase transition with the well-known phase decomposition in BIM-model [2,3].
3. Classical OCP of charged hard and soft spheres.
4. Classical point charges (+background) with the repulsion $\sim 1/r^s$ ($s \leq 2$), and Hard Spheres with the repulsion $\sim 1/r^s$ ($s \leq 3$).